**Title:** **A Novel Brain–Computer Interface Architecture: The Brain–Muscle–Hand Interface for replicating the motor pathway**

**Authors:** Sun Ye[1, 2, †], Zuo Cuiming[1, 3, †], Zhang Rui[1, 2, †], Shi Bin[4], Pang Yajing[1, 2],Gao Lingyun[1, 2], Zhao Bowei[1], Wang Jing[3], Yao Dezhong[5, 6, 1, 2, *], Liu Gang[1, 2, *]

[1] School of Electrical and Information Engineering, Zhengzhou University; Zhengzhou, 450001, China

[2] Henan Provincial Key Laboratory of Brain Science and Brain-Computer Interface Technology; Henan, China

[3] School of Mechanical Engineering, Xi'an Jiaotong University, Xi'an, 710049, China

[4] Rocket Force University of Engineering; Xi'an, 710025, Shaanxi, China

[5] Clinical Hospital of Chengdu Brain Science Institute, MOE-Key Lab for NeuroInformation, Brain-Apparatus Communication Institute, University of Electronic Science and Technology of China, Chengdu, China

[6] Research Unit of NeuroInformation 2019RU035, Chinese Academy of Medical Sciences, Chengdu, 611731, China

* Liu Gang. Email: gangliu_@zzu.edu.cn；Yao Dezhong. Email: dyao@uestc.edu.cn.

**Abstract:** Myoelectric interfaces enable intuitive and natural control by decoding residual muscle activity, providing an effective pathway for motor restoration in individuals with preserved musculature. However, in patients with severe muscular atrophy or high-level spinal cord injury, the absence of reliable muscle activity renders myoelectric control infeasible. In such cases, motor brain–computer interfaces (BCIs) offer an alternative route. However, conventional brain–computer interface systems rely mainly on noisy cortical signals and classification-based decoding algorithms, which often result in low signal fidelity, limited controllability, and unstable real-time performance. Inspired by the motor pathway—an evolutionarily optimized system that filters, integrates, and transmits motor commands from the brain to the muscles—this study proposes the Brain–Muscle–Hand Interface (BMHI). BMHI decodes cortical EEG signals to reconstruct muscle-level EMG activity, functionally substituting for the muscles and enabling regression-based, continuous, and natural control via a myoelectric interface. To validate this architecture, we performed offline verification, comparative analysis, and online control experiments. Results demonstrate that: (1) the BMHI achieves a prediction accuracy of 0.79; (2) compared with conventional end-to-end brain–hand interfaces, it reduces training time by approximately eighteenfold while improving decoding accuracy; and (3) in online operation, the BMHI enables stable and efficient manipulation of both a virtual hand and a robotic arm. Compared with conventional BCIs, the BMHI, by replicating the motor pathway, enables continuous, stable, and naturally intuitive control.

**One-Sentence Summary:** This paper proposes a novel motor BCI—the BMHI. The proposed approach replicates the motor pathway to enable real-time reconstruction of muscle activity. The reconstructed EMG signals derived from EEG are further translated into control commands, allowing the system to retain the advantages of regression-based myoelectric interfaces that provide natural and continuous control. Compared with traditional classification-based motor

BCIs, the BMHI demonstrates superior performance in motor control, as its regression-based architecture enables continuous, natural, and intuitive movement generation.

**Main Text:**

## INTRODUCTION

Globally, tens of millions of individuals suffer from severe motor impairments due to limb amputation, muscular atrophy, or high-level spinal cord injury[1-4]. These conditions, resulting from trauma, vascular disease, neurodegenerative disorders, or congenital defects, not only lead to profound loss of movement capability but also impose long-term psychological and social burdens, severely diminishing quality of life[5, 6]. While individuals with lower-level amputations or partial muscle retention can rely on conventional myoelectric interfaces or exoskeletal control systems for relatively natural motion, those with extensive muscular atrophy or complete paralysis often lose the ability to generate reliable EMG. Consequently, traditional Muscle interfaces-based approaches become ineffective[7-9]. For these patients—who lack accessible muscular output—motor BCIs represent the only viable pathway to restore voluntary control[10, 11].

However, current BCIs, such as MI-BCIs[12-15], generally adopt a "direct brain-to-device" control paradigm. These systems largely rely on the recognition of a limited set of electroencephalographic patterns, resulting in discrete command outputs rather than continuous, natural control signals. This approach bypasses the complex and highly efficient motor neuro pathways refined over hundreds of millions of years of evolution[16, 17]. These pathways do not merely serve as conduits for signal transmission; they also function as biological signal processors, optimizing, filtering, and integrating cortical outputs to ultimately produce refined and fluid motor commands [18-20]. In contrast, scalp-recorded EEG signals are inherently noisy, highly blended, and susceptible to interference from non-motor brain activity[21, 22]. During natural movement, the human nervous system employs multi-level coordination to apply a 'biological filter' to high-level intent, translating it into precise, continuous task-relevant muscle activity. Traditional BCIs overlook this naturally evolved 'filtering mechanism', attempting instead to reconstruct control signals directly from raw cortical activity. Consequently, they often exhibit low signal fidelity, limited control dimensions, poor robustness, and non-intuitive feedback, which restricts their reliability and user acceptance in practical applications[23, 24].

To address these challenges, this paper proposes the BMHI framework, which is centered on replicating the motor pathway. It first decodes EEG signals to extract motor intent, then uses this information to reconstruct muscle signals in real time, and finally integrates this with EMG control paradigms to form a complete control pathway. This approach preserves the key advantages of EMG-based control—namely, its biologically constrained, regression-based framework that enables natural and continuous motor control. Thereby, it achieves a control experience that is more continuous and natural than what is possible with traditional classification-based motor BCIs, which offers a novel approach for providing natural and adaptive motor control to individuals with muscle atrophy, spinal cord injuries, and severe limb loss.

This study is anchored in advancing the core technology of brain–computer interfaces. It focuses on establishing and validating a feasible framework based on the brain–muscle mapping mechanism, aiming to overcome the limitations of conventional brain–computer interfaces from the perspective of neural information decoding and to achieve more efficient modelling and transmission of motor intentions. Using healthy participants, we experimentally test whether the artificial motor pathway replicated by BMHI can functionally substitute for the biological motor pathway to control a prosthetic or robotic hand and execute movements.

The main contributions of this paper include:

1. This paper first proposes the Brain–Muscle–Hand Interface (BMHI), which replicates the motor pathway from the brain to the muscles, enabling real-time reconstruction of muscle activity. The reconstructed muscle activities through this artificial motor pathway can be seamlessly integrated with myoelectric interfaces, enabling the BMHI to inherit the intrinsic advantages of myoelectric control. Consequently, it achieves physiologically constrained, regression-based, and natural continuous control, while overcoming the limitations of traditional classification-based BCIs in terms of decoding accuracy and movement naturalness.
2. This paper develops a real-time neural interface system that achieves continuous motor control by replicating the motor pathway and forms a closed-loop, allowing users to actively modulate cortical activity through real-time feedback. This capability for stable and natural control highlights the system's promise for advanced neurorehabilitation and intuitive user applications.
3. This paper proposes a trend-based error loss function that captures temporal evolution in neural time-series data, enabling the model to learn trend-level motion intentions rather than discrete frame-wise fluctuations. This algorithmic innovation enhances the robustness and controllability of online decoding, supporting smooth and continuous brain-driven motor control.

This paper is structured as follows: Section 2 presents the experimental results; Section 3 discusses the significance of the findings, along with the summary and future directions; Section 4 details the method's design and implementation.

## RESULTS

In healthy individuals, movement is generated through the motor pathway: the brain produces motor intentions, which are then filtered and integrated along descending neural circuits[18, 25]. These processed and refined signals are transmitted to the muscles, where contraction and relaxation drive tendon motion, ultimately leading to joint movement.

The objective of the proposed BMHI is to replicate this natural motor pathway by artificially reconstructing the transmission of neural signals from the brain to the muscles. In contrast to traditional classification-based brain-computer interfaces, which directly map cortical activity to movements, the BMHI decodes electroencephalography signals to reconstruct corresponding electromyography signals. This approach restores the lost neuromuscular connection in individuals lacking functional musculature. The reconstructed muscle signals serve as an artificial electromyography that can integrate with existing myoelectric systems, thereby inheriting their inherent advantages—namely, natural, continuous, and regression-based control.

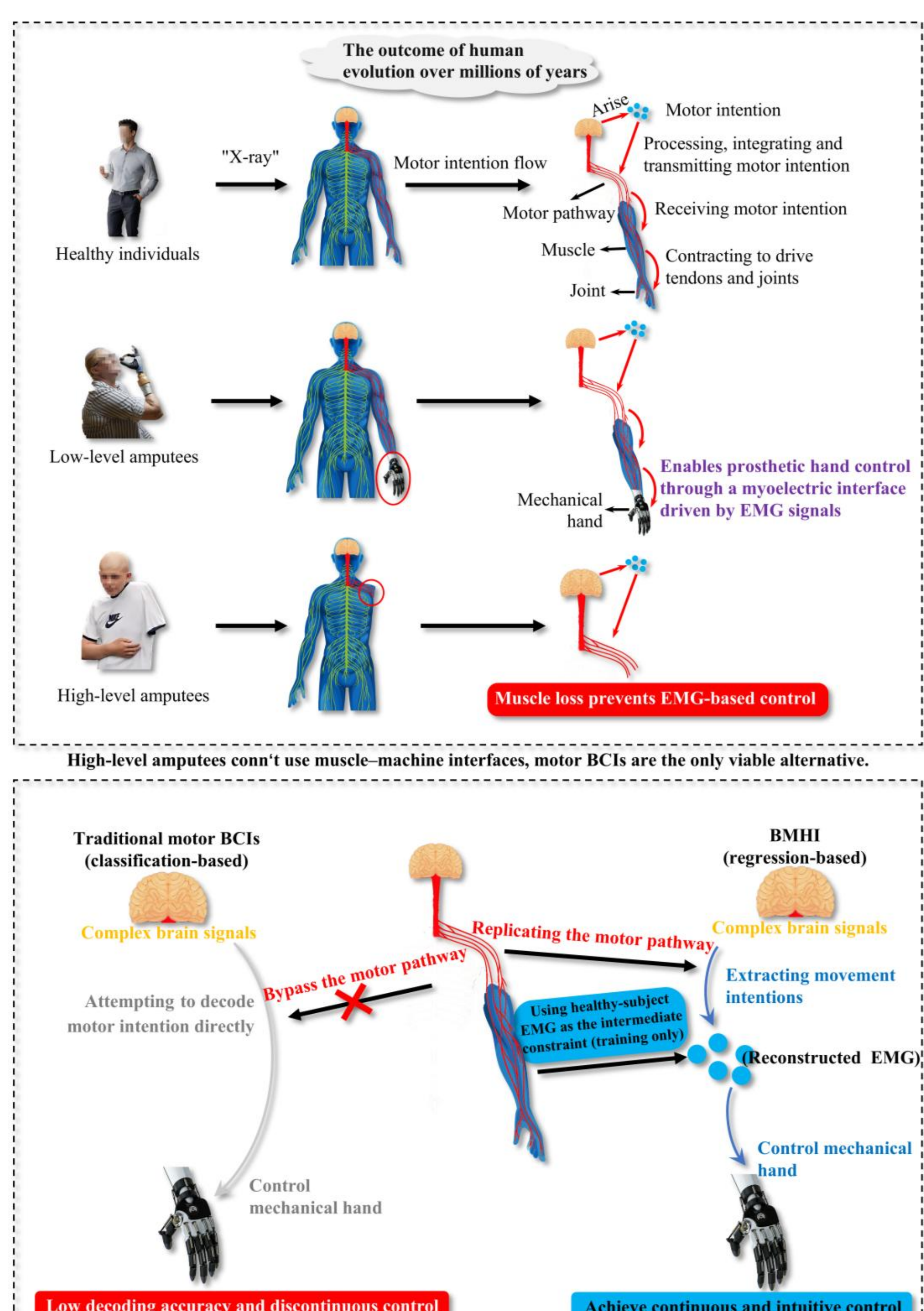

**Fig. 1. Brain–Muscle–Hand Interface (BMHI): Replicating the Human Motor Pathway.** Schematic illustration of motor intention transmission in healthy individuals and amputees. In healthy subjects, motor intentions generated in the brain are processed and transmitted through motor nerves to muscles, driving joint motion. Low-level amputees can control prosthetic hands through myoelectric interfaces by decoding residual EMG signals. However, high-level amputees lose muscular pathways, preventing EMG-based control. Traditional motor BCIs attempt to bypass

the motor pathway and directly decode motor intentions from complex brain signals, resulting in low decoding accuracy and discontinuous control. The proposed BMHI reconstructs EMG by replicating the human motor pathway, thereby achieving more precise and intuitive control of prosthetic hands.

## Offline Experiment

### *Can the EMG signals reconstructed by the BMHI-replicated motor pathway truly reflect the EMG signals generated in the human motor pathway?*

To validate whether the EMG signals reconstructed by the replicated artificial motor pathway in BMHI accurately represent those generated by the human motor system, an offline experiment was conducted. Thirty-five healthy participants were recruited, all right-handed, without neurological or muscular disorders, and with written informed consent. The experimental protocol was approved by the Ethics Committee of Zhengzhou University (Approval No. 62303423) and complied with the Declaration of Helsinki and relevant Chinese ethical standards.

EEG signals were recorded using a 16-channel EEG system, while 12 surface electrodes synchronously captured forearm EMG activity and 10 pressure sensors monitored finger forces (see Fig. 2). Each participant performed six predefined hand movements, repeated ten times, yielding sixty synchronized EEG–EMG–force samples per subject.

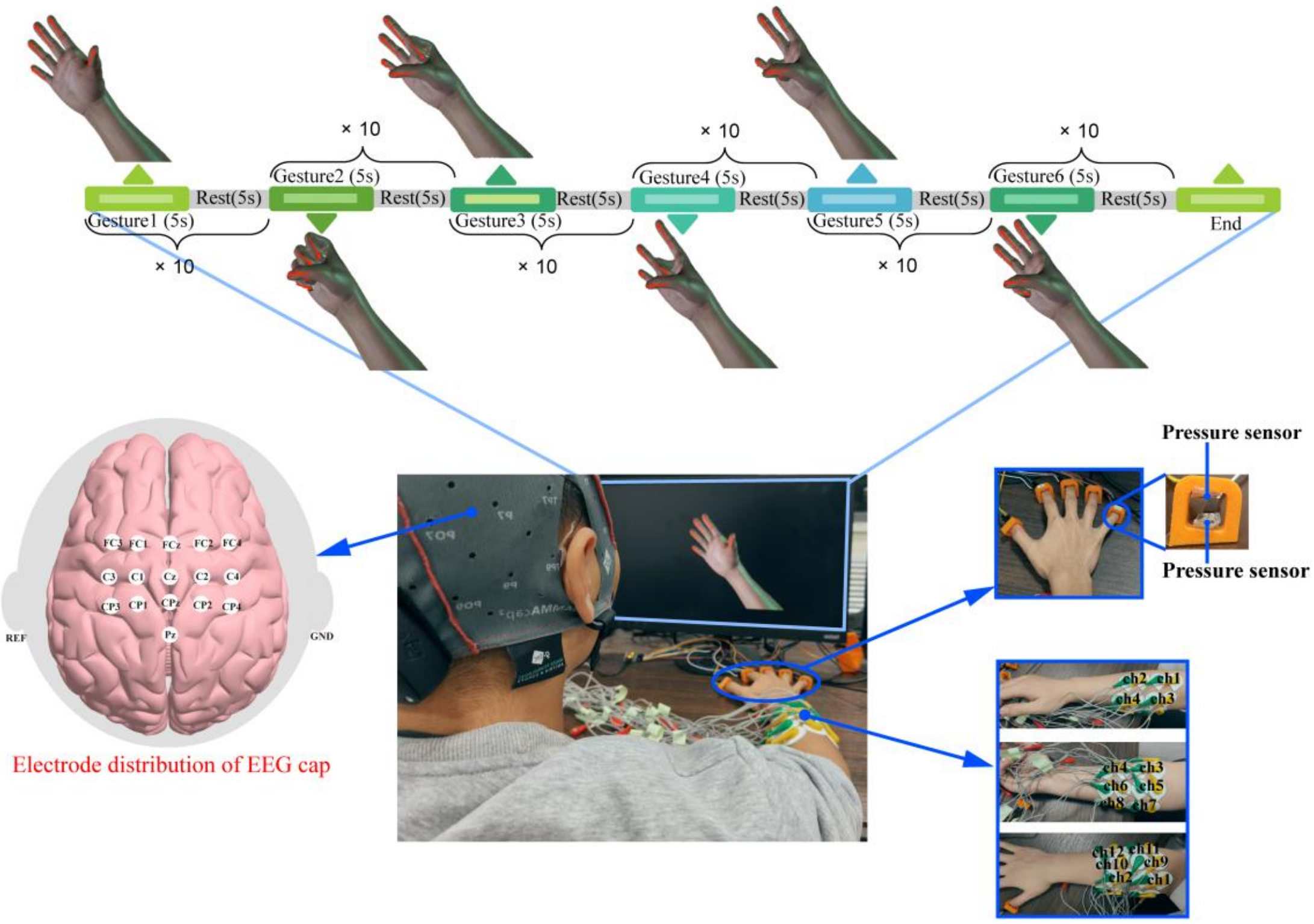

**Fig. 2. Experimental paradigm.**

The decoding performance of BMHI was quantitatively evaluated using the Pearson correlation coefficient (PCC) between the reconstructed and actual EMG signals. A representative example is shown in Fig. 3A, where the EMG signals reconstructed by the BMHI exhibit a high degree of waveform consistency with the actual EMG recordings. At the population level, PCC values were computed for all 12 EMG channels across 35 participants. The mean PCC of each participant and the average of the maximum PCCs across participants were subsequently derived. As illustrated in Fig. 3B, the overall mean PCC was 0.16, the average of the maximum PCCs reached 0.35, and the highest PCC observed across all participants was 0.79. These findings demonstrate that the EMG signals reconstructed by the BMHI-recreated motor pathway are able to reflect those generated within the motor pathway. Notably, the “average of the maximum PCC” represents the theoretical performance level attainable by the model under optimal channel conditions.

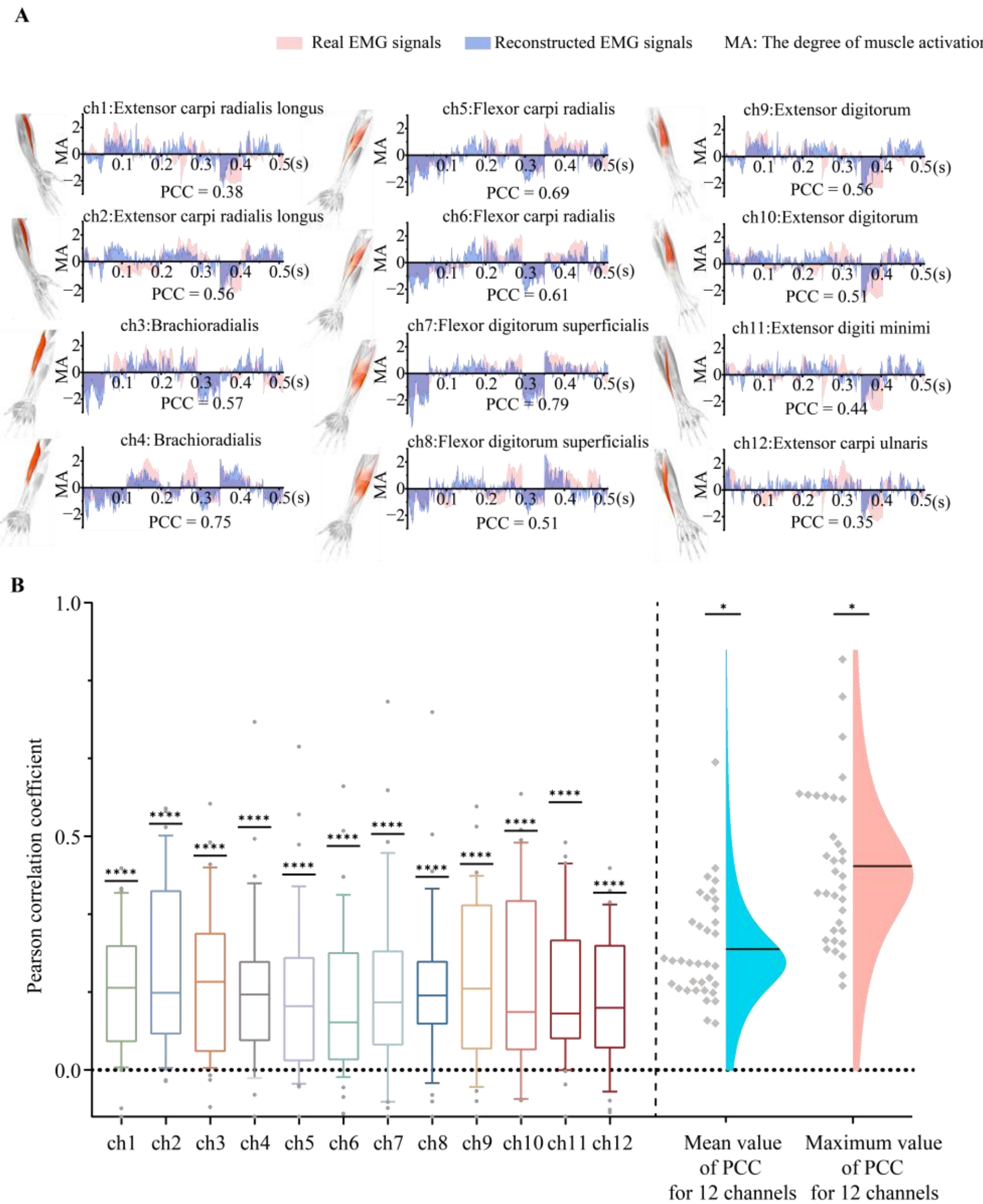

**Fig. 3. Comparison between reconstructed and measured EMG signals and statistical validation of the BMHI performance.** (A) Reconstructed EMG signals vs. real EMG signals. The EMG signal has 12 channels; PCC denotes Pearson correlation coefficient, and muscle activation degree is measured by integrated electromyography; (B)Coefficients between reconstructed and actual 12-channel values of EMG signals in 35 subjects. A one-sample t-test was performed on the correlation coefficients of 12-channel EMG signals compared to 0 (no correlation). Results showed all channels had coefficients significantly above 0 ($p < 0.001$),

indicating a correlation between reconstructed and actual EMG signals. This confirms the efficacy of the brain-muscle interface method;

The one-sample *t*-test further confirmed the statistical significance of the correlations, comparing PCC values against zero (representing no correlation, i.e., failure to extract motor intention). The results revealed that both the mean and maximum PCC values across all twelve channels were significantly greater than zero ($p < 0.05$), confirming that the reconstructed EMG signals reflect genuine motor-related activity rather than random noise. Collectively, these results confirm that the artificial motor pathway replicated by BMHI can effectively reconstruct the muscle activities in the motor pathway.

***Can the artificial motor pathway replicated by BMHI faithfully reproduce hand movements driven by the human motor pathway?***

To further verify the ability of the artificial motor pathway replicated in BMHI to faithfully reproduce motor pathway–driven hand movements, the reconstructed EMG signals were integrated into a myoelectric interface to predict finger forces. As illustrated in Fig. 4A, the predicted finger forces showed a degree with the actual recorded forces. Population level analysis indicated that the overall mean PCC across the five finger channels was 0.24, while the average of the maximum PCC values across all participants reached 0.57, with a global maximum of 0.80 (see Fig. 4B).

The one-sample t-test results further confirmed that the mean PCC for the five finger channels was significantly above 0.3($p < 0.0001$), and the maximum PCC was significantly above 0.47 ($p < 0.0001$), suggesting that the EMG signals reconstructed by the BMHI were not only statistically valid but also contained sufficient motor information for downstream movement prediction. These findings demonstrate that the artificial motor pathway replicated by BMHI can reproduce hand movements driven by the motor pathway, verifying its feasibility in transmitting motor intentions and achieving functional substitution.

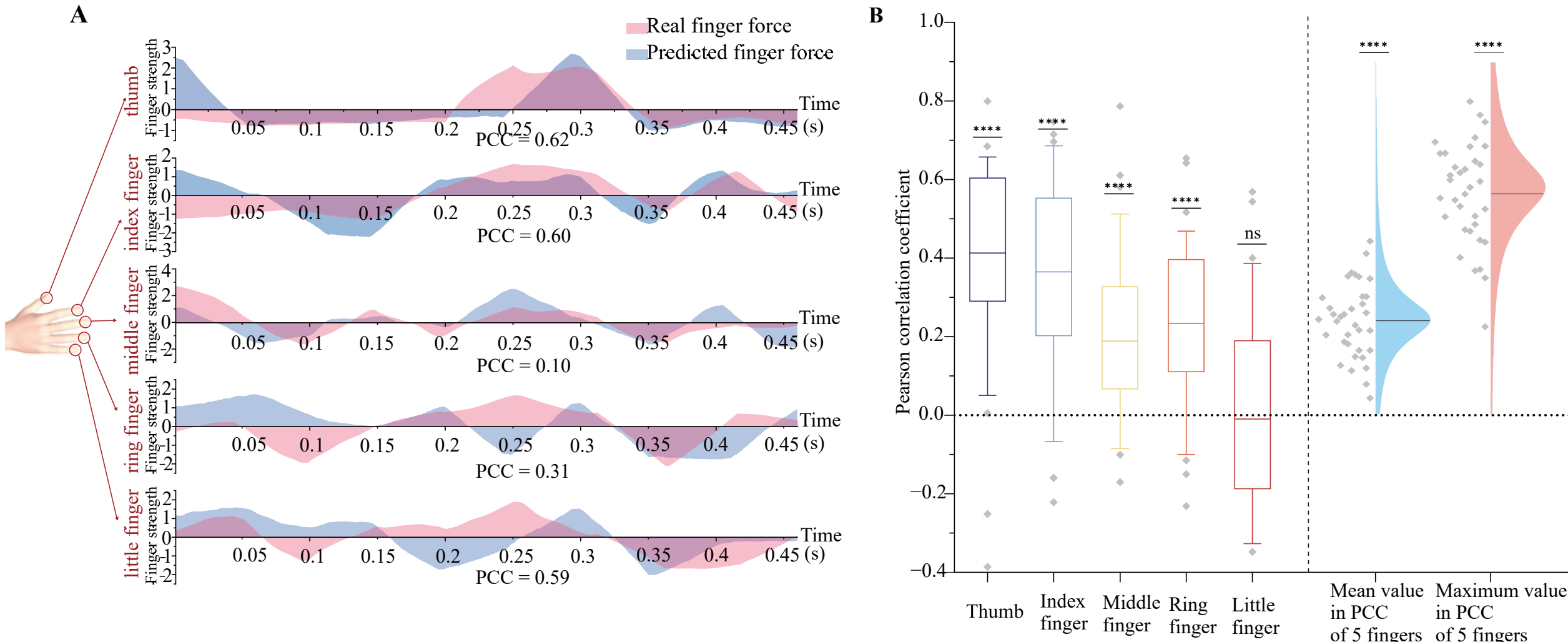

**Fig. 4. Comparison between predicted and measured finger forces and statistical validation of the BMHI performance.** (A) Predicted versus measured forces for the five fingers; (B)Statistical analysis of correlation coefficients for predicted versus actual five-finger strength measurements in 35 participants. A one-sample t test was performed for the correlation coefficient of 5 finger forces with 0 (no correlation). The results showed that, except for the little finger, the

correlation coefficients of the other four fingers were significantly greater than 0 ($p < 0.001$), indicating a correspondence between predicted and actual motor outputs. Importantly, the weak correlation observed in the little finger ($p > 0.05$) is consistent with its biomechanical role as a non-independent digit—its motion is passively coupled with adjacent fingers through shared tendon linkages, which inherently limits the separability of its isolated force patterns[26]. This finding suggests that the BMHI not only achieves accurate decoding of individual finger forces but also faithfully reflects the intrinsic coupling structure embedded in the human Motor pathway. Despite partial biomechanical coupling between the ring finger and neighboring digits, the high correlation coefficients of the thumb, index, middle, and ring fingers ($p < 0.001$) further confirm that BMHI captures the synergistic activation patterns characteristic of natural hand movement.

**Controlled experiment**

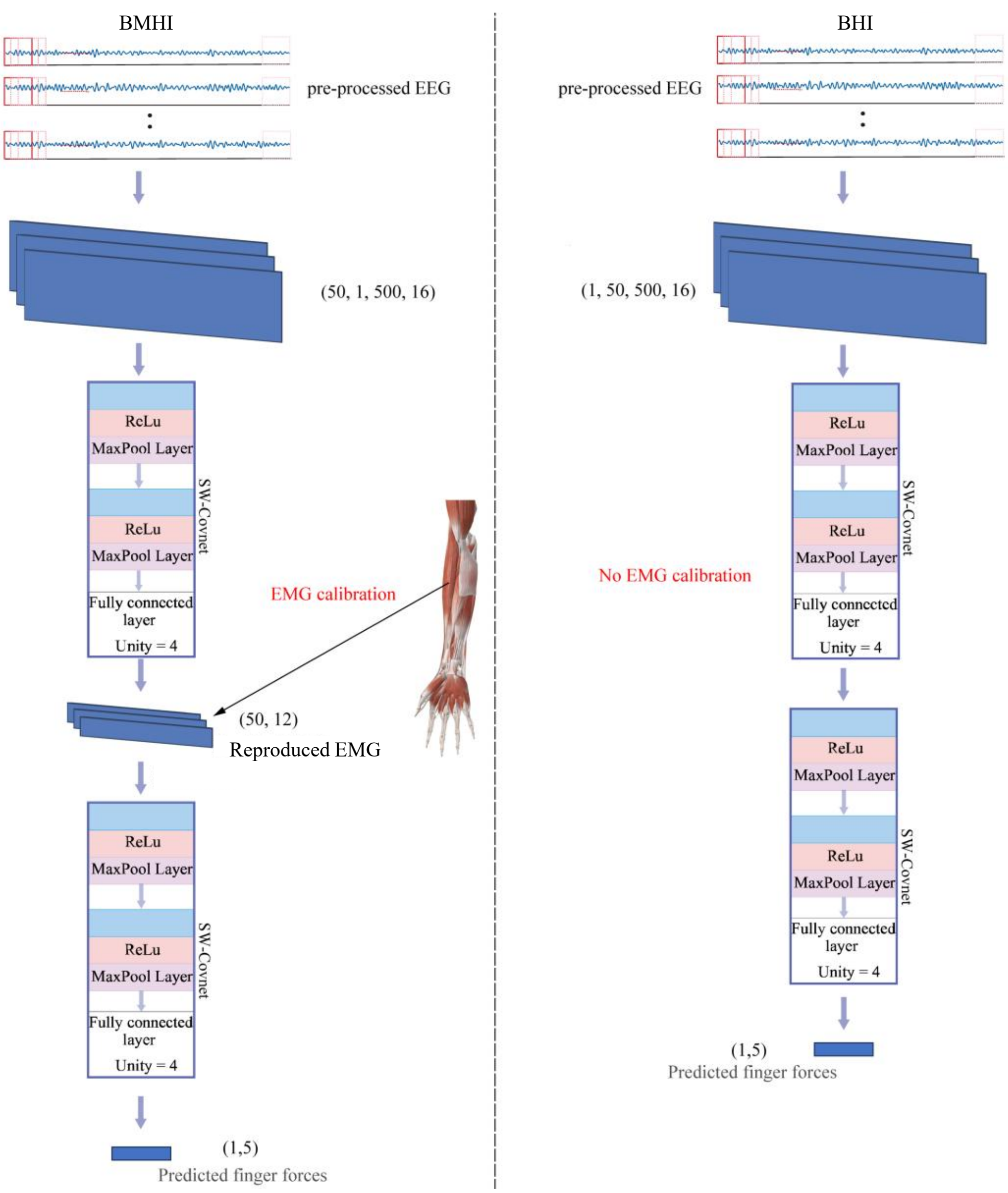

**Fig. 5. Comparison of BMHI and BHI.** The BMHI replicates the motor pathway, decodes EEG to reconstruct EMG activity, and thereby drives the prosthetic hand via a myoelectric interface. In contrast, the Brain–Hand interface (BHI) bypasses the physiological pathway and directly decodes EEG for control.

To evaluate the practicality and performance of the BMHI method, a comparative experiment was conducted under identical model structures and parameter settings. The experimental group employed the BMHI approach, which decodes cortical EEG signals by reconstructing the Motor pathway to reproduce muscle-level EMG activity. These reconstructed EMG signals were then used by a myoelectric interface for continuous and natural finger force prediction. In contrast, the control group adopted a conventional Brain-Hand interface (BHI) approach. This method bypasses the underlying neurophysiological pathways and directly predicts finger forces from EEG signals.

Both models were evaluated by comparing the Pearson correlation coefficients (PCCs) between the predicted and the actual finger forces. Figure 6A presents the statistical results of the prediction accuracy for the two methods. To further assess model efficiency, the training time for both approaches was also recorded (see Fig. 6B).

Experimental results demonstrated that the BMHI method significantly outperformed the traditional BHI approach in decoding performance, with a training speed approximately eighteen times faster. These findings validate the effectiveness of the BMHI in replicating human motor pathway. By reconstructing muscle-level EMG signals through the replication of the Motor pathway, BMHI enhances the physiological fidelity of motor decoding and enables continuous, natural, and efficient extraction of motor intentions.

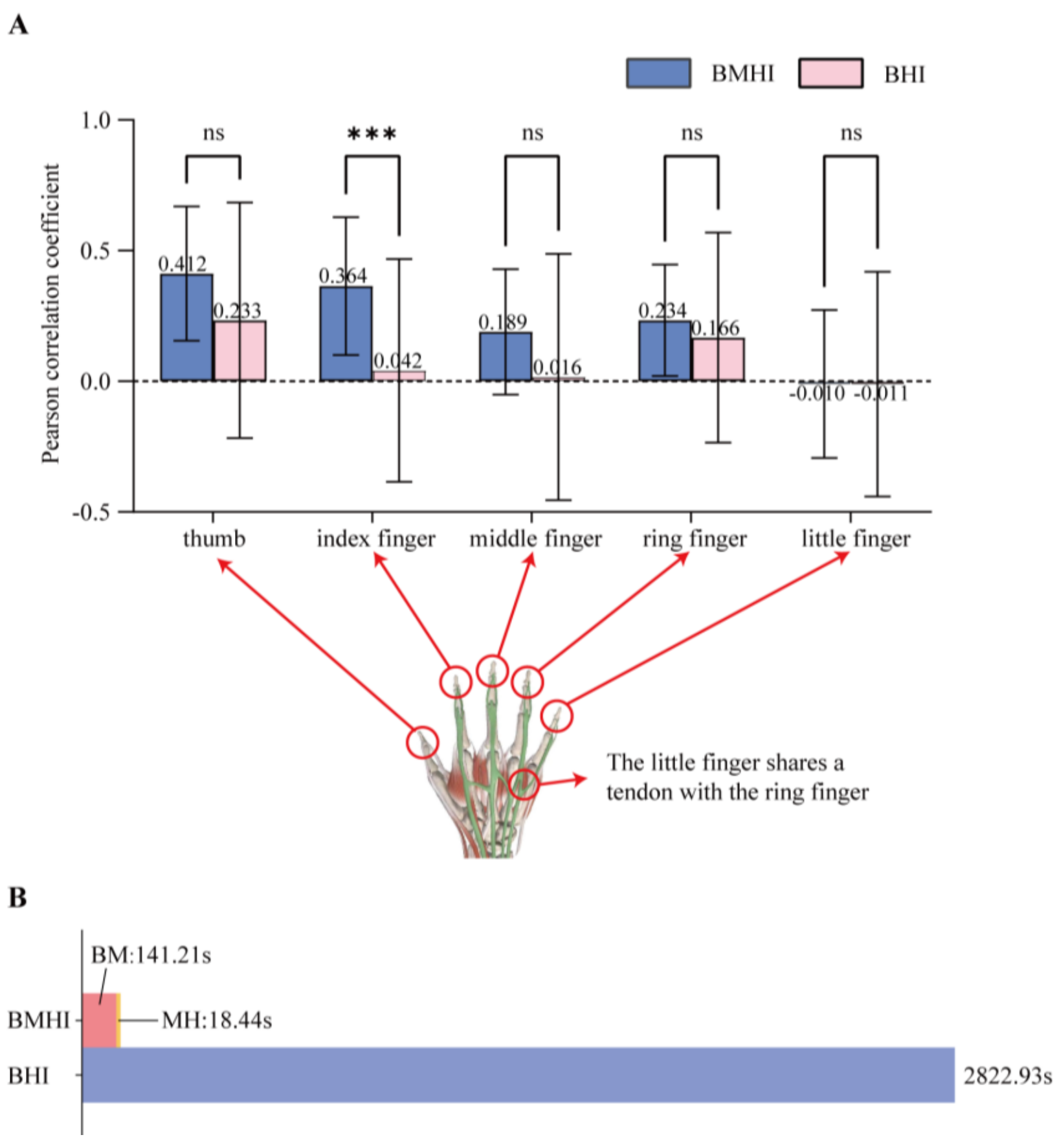

**Fig. 6 Comparison results of BMHI and BHI.** (A)BMHI and BHI performance comparison to evaluate BMHI and BHI performance by comparing the Pearson correlation coefficient between predicted finger force and true finger force; (B)Comparison of training time between BMHI and BHI models.

**Online experiment**

Considering the coupling effects observed in multi-finger movements during the offline experiments, the online experiments were designed with two representative actions— "hand opening" and "fist clenching" —to minimize interference and more precisely evaluate the real-time control performance of the BMHI.

***Whether the motor pathway replicated by BMHI is capable of accurate real-time control?***

To investigate whether the artificial motor pathway replicated by BMHI is capable of accurate real-time control, we designed an online brain-controlled experiment comprising two parts: a virtual hand task and a physical robotic hand task. These were used to evaluate the system's real-time decoding performance and control responsiveness in practical operational settings. For each participant, a personalized BMHI model was trained using their offline data. During the online phase, EEG signals were decoded in real time to reconstruct muscle-level EMG activity, which was then transmitted through the myoelectric interface and converted into corresponding finger force outputs to drive the target devices (see Fig. 7). Three participants were recruited for the online experiments. Prior to the online session, each participant performed two motor tasks— "hand opening" and "fist clenching"—repeated 60 times to collect individual offline EEG–EMG datasets for model training and calibration.

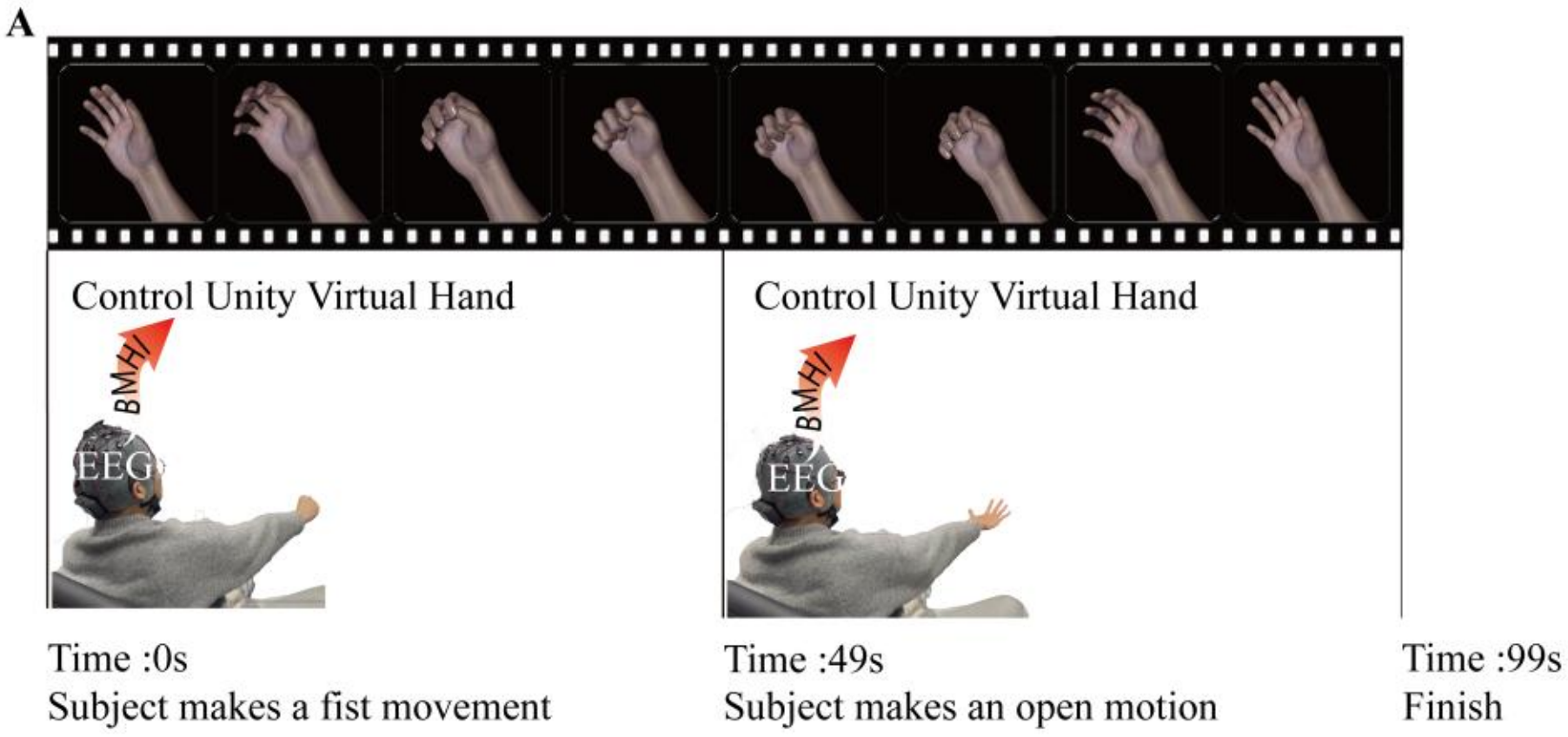

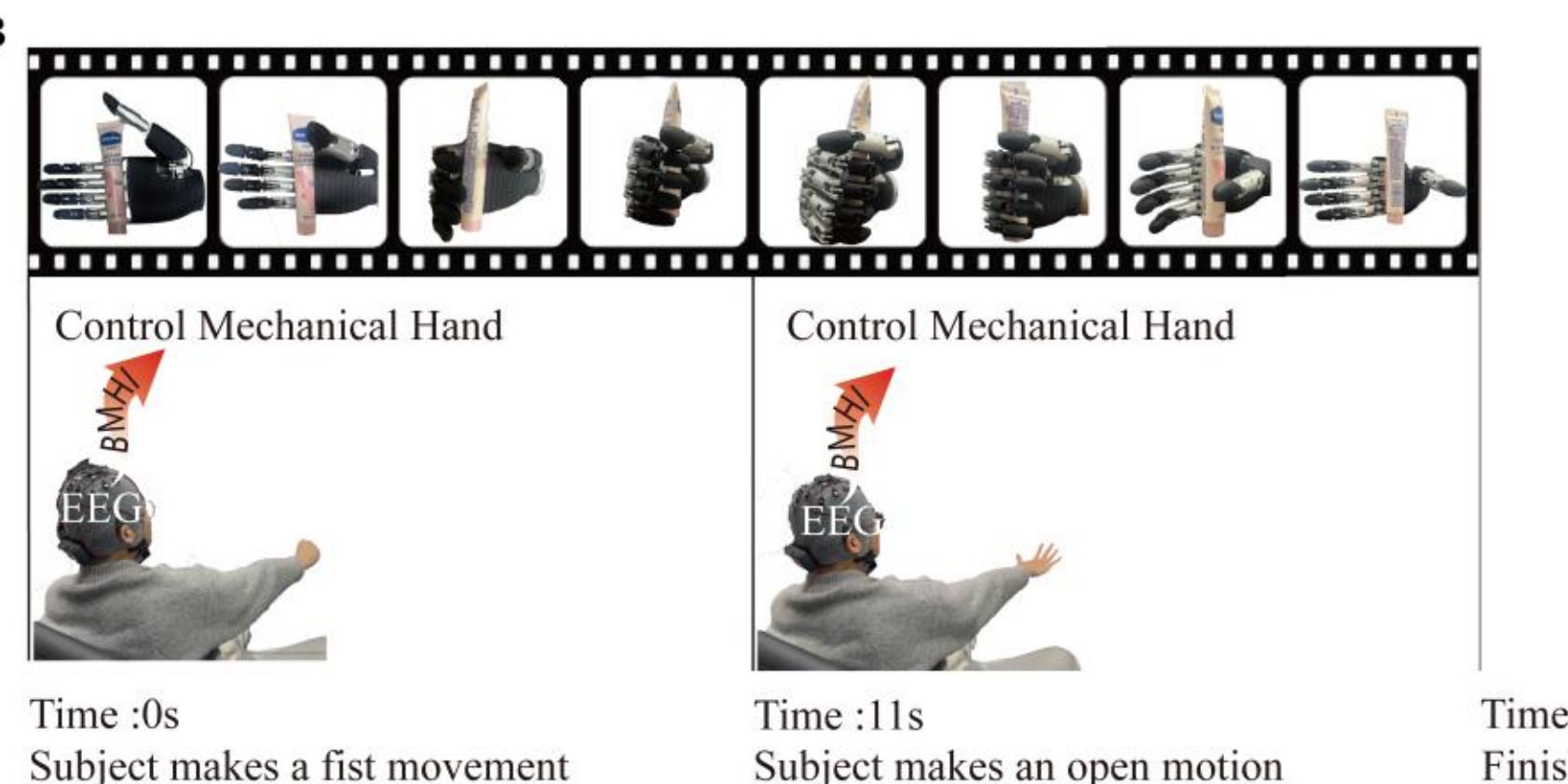

**Fig. 7. Subjects control virtual/ mechanical hand movements.** (A) Subject controls unity virtual hand movement; (B) Subject controls manipulator movement.

***Experiment I: Virtual Hand Control***

In the virtual hand task, the finger forces predicted by the BMHI were mapped in real time to control signals for a Unity-based virtual hand, enabling brain-controlled grasping (Fig. 7A). Results showed that the average completion times for the grasping and opening tasks were 27.67s and 58.67s, respectively (see Table 1). Under closed-loop feedback conditions, participants were able to voluntarily modulate their brain activity based on visual feedback, achieving continuous movement control (see Supplementary Videos S1–S4). These results demonstrate that the brain–muscle pathway constructed by the BMHI enables stable intention transmission and control execution in real-time feedback environments, exhibiting features of naturalistic control.

**Table. 1. Time taken by participants to complete the Unity virtual hand online experiment.**

| Participants | Time required to make a fist | Time to open fingers |
|---|---|---|
| Participant 1 | 14s | 96s |
| Participant 2 | 20s | 30s |
| Participant 3 | 49s | 50s |

***Experiment II: Robotic Hand Control***

To further validate its practicality, a robotic hand control experiment was conducted using the same mechanism (see Supplementary Videos S5–S9). The BMHI decoded EEG signals and reconstructed EMG activity in real time, which was subsequently integrated into the myoelectric–mechanical interface to infer finger force outputs, thereby converting them into control commands to drive a physical robotic hand (Fig. 7B). In a multi-object grasping task simulating real-world scenarios, participants achieved average completion times of 12s for grasping and 13s for releasing (see Table 2). These findings indicate that the BMHI maintained high responsiveness and control consistency in real-world interaction contexts, confirming the system's stability and generalizability for practical brain-controlled applications.

**Table.2. Time taken by participants to complete the mechanical hand online experiment.**

| Participants | Time required to make a fist | Time to open fingers |
|---|---|---|
| Participant 1 | 19s | 23s |
| Participant 2 | 6s | 6s |
| Participant 3 | 11s | 10s |

The results indicate that BMHI reconstructs the motor pathway to decode cortical EEG signals and reproduce EMG activity, which is then transmitted through a myoelectric interface to generate force outputs for natural finger-movement control. This regression-based mechanism enables continuous and natural motor control, bridging neural decoding and motor execution through an artificial motor pathway. These properties validate BMHI's potential for real-time closed-loop BCI applications, including cortically controlled prosthetics, motor rehabilitation for paralysis, and adaptive human–robot collaboration.

**Model Supplement Experiment**

To verify the effectiveness of the proposed trend-based error loss function in time-series neural decoding, a comparative experiment was conducted using offline data from thirty-five participants. The experiment was performed under identical data partitions and model structures. In the control group, only the mean squared error (MSE) loss was applied, whereas the experimental group employed a composite loss function that incorporated an additional trend-based error term on top of the MSE. The experimental procedure and structure of the loss function are illustrated in Fig. 8.

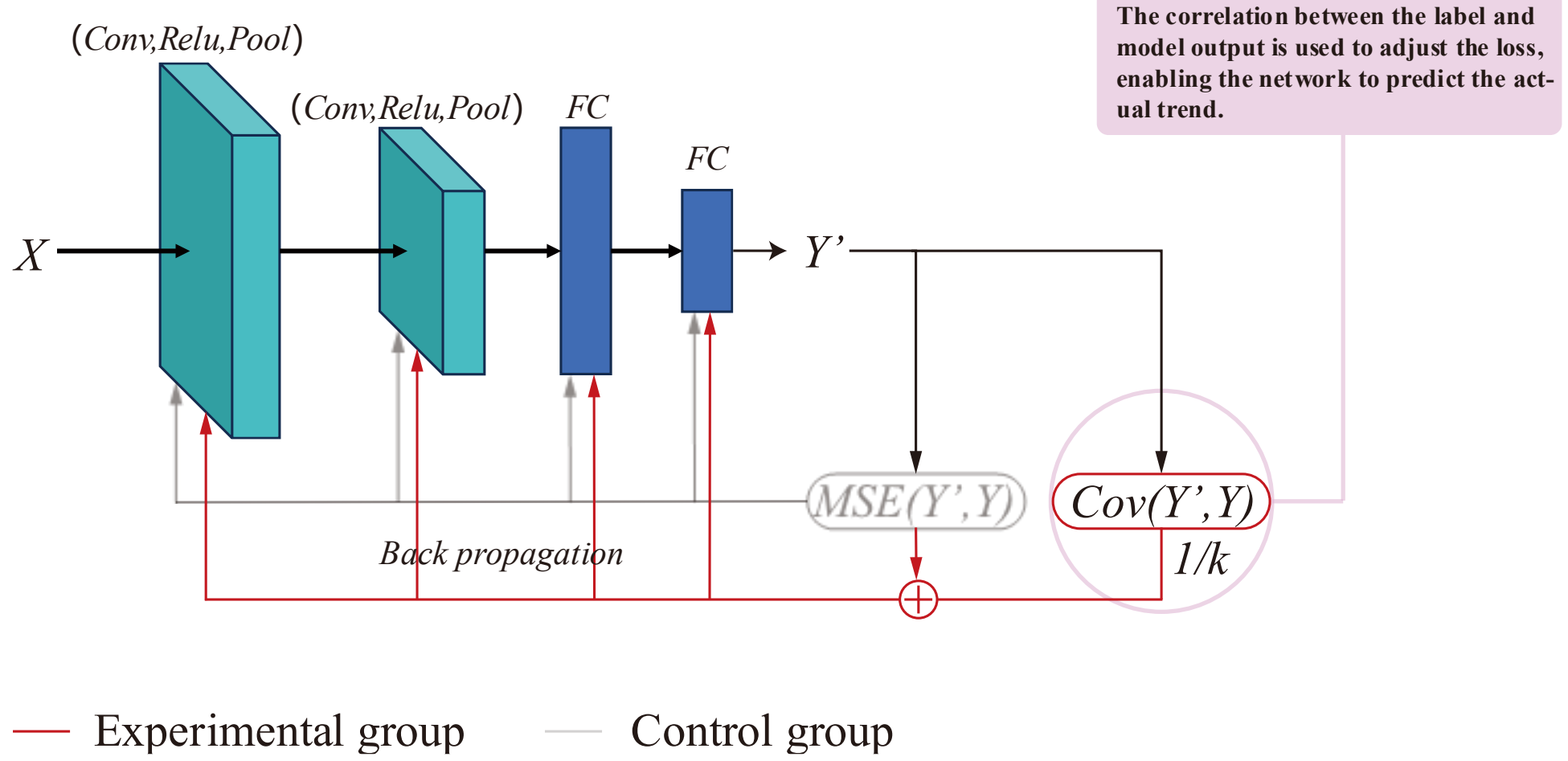

**Fig. 8. Experimental group vs. control group setup diagram.**

To assess performance, the Pearson correlation coefficient (PCC) was used to quantify the temporal consistency between the predicted and actual signals. Specifically, the mean PCC value across all channels was first calculated for each participant, followed by averaging across all thirty-five participants to reflect group-level performance. To emphasize the impact of the loss function on temporal trend reconstruction, the analysis focused on mean PCC values rather than maximum PCC values, thereby avoiding bias from isolated channels with high correlations. The results are presented in Fig. 9.

The findings show that, after introducing the trend-based error term, the model achieved higher mean PCC values for most participants, indicating an overall improvement in decoding performance compared with the model trained using only MSE loss. This demonstrates that the trend constraint enhances the model's ability to learn temporal variation patterns within the signals, thereby improving the temporal consistency and stability of brain–muscle decoding. Overall, the trend-based error loss function effectively increased the decoding accuracy of the BMHI, confirming the validity of the proposed approach.

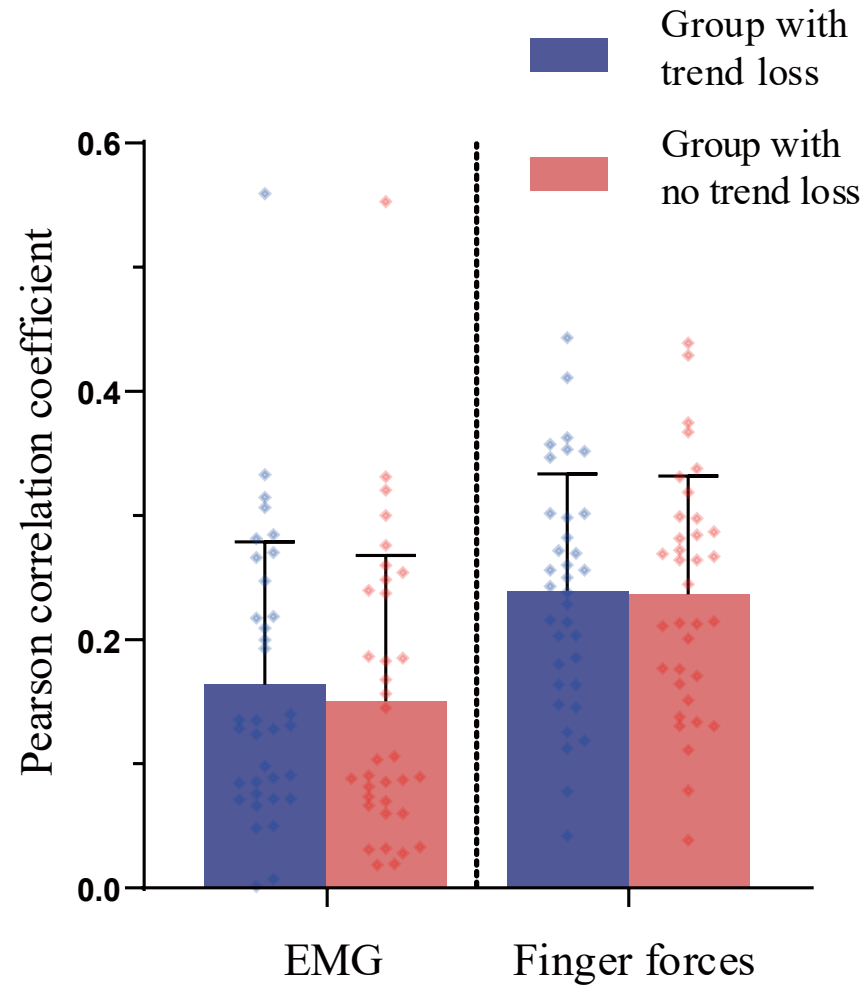

**Fig. 9. Statistical of PCC between reconstructed and true values of EMG and gestures in comparison experiments.**

## DISCUSSION

The BMHI proposed in this study aims to replicate the motor pathway. Unlike conventional motor BCIs that directly map cortical activity to motor outputs—bypassing the intricately evolved neuromuscular pathways refined over hundreds of millions of years—BMHI reconstructs this biological control hierarchy to decode cortical EEG signals and reproduce EMG activity. In doing so, it restores the intermediate “brain–to–muscle” stage that is fundamental to natural motor regulation. The reconstructed EMG signals can then be transmitted to myoelectric interfaces, thereby inheriting their intrinsic advantages of regression-based, continuous, and intuitive control (see Fig. 10).

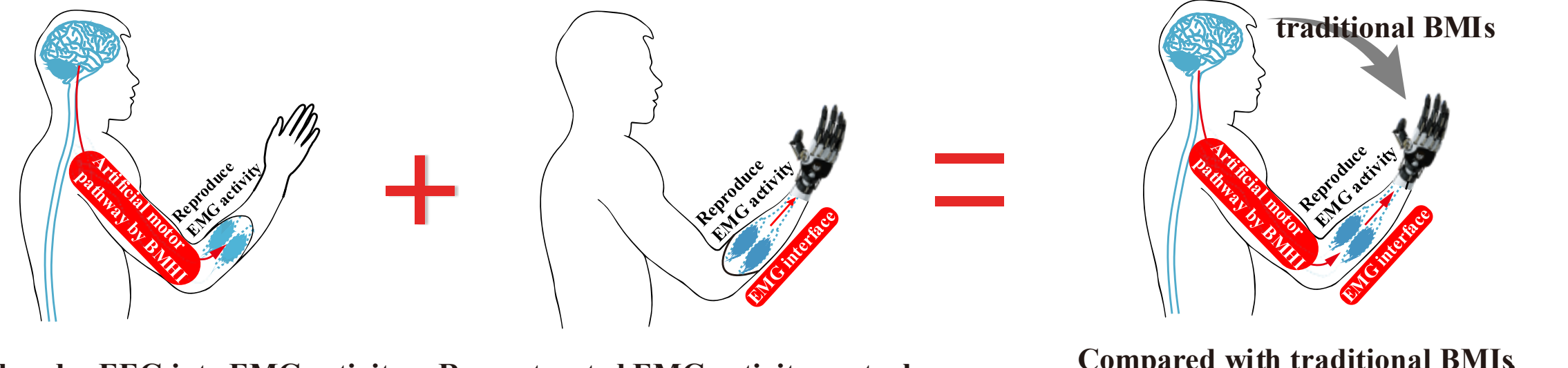

**Fig. 10 Schematic of the BMHI system.** Unlike conventional BCIs that bypass neuromuscular pathways, the BMHI decodes EEG to reproduce EMG activity, thereby replicate the biological "brain-to-muscle" control hierarchy. This reconstructed EMG is then utilized by a myoelectric interface for regression-based, continuous, and intuitive prosthetic hand control.

Previous studies have demonstrated that, during spontaneous or intentional limb movements, EEG and EMG signals exhibit frequency-domain coupling, reflecting an intrinsic connection between cortical activity and muscle activation[27-29]. This neuro–muscular co-activation provides an essential physiological foundation for the design of the BMHI. Leveraging the intrinsic coupling between cortical and muscular activity, BMHI reconstructs muscle activation from EEG, thereby replicating the natural motor pathway and restoring the physiological “brain-to-muscle” link fundamental to biological movement control.

From the perspective of neural information decoding, this mechanism overcomes the long-standing limitations of conventional brain–computer interfaces in modelling and transmitting motor intentions. Traditional systems typically bypass the natural neural regulation pathways by directly generating motor commands from EEG signals, thereby neglecting the body’s intrinsic mechanisms of integration and filtering during signal transmission. In contrast, BMHI integrates the intrinsic modulation and transmission mechanisms of the motor pathway into the decoding process, allowing neural signals to be progressively refined before being transformed into motor commands.

To validate the effectiveness and physiological plausibility of the BMHI framework, this study first systematically evaluated its decoding performance for movement intention through offline experiments, with results from 35 subjects demonstrating that BMHI performed well in both the extraction and usability of motor intention: in terms of intention extraction, BMHI could reconstruct corresponding muscle activity from EEG signals, and the reconstructed EMG signals showed correlation with the recorded signals, indicating its capability to effectively capture movement intention information from the brain; in terms of intention usability, the predicted finger movement signals also quite correlation with the actual movement signals, proving that the decoded motor intention could be effectively translated into actual finger movements; further analysis revealed that the Pearson correlation coefficients for the thumb and index finger were significantly higher than those for the middle, ring, and little fingers, a phenomenon consistent with existing studies on the physiological coupling characteristics of hand movements, where the middle, ring, and little fingers tend to exhibit stronger muscular synergy and neural coupling, with the little finger showing particularly high dependency during movement execution[30-33]; based on this observation, subsequent online experiments focused on two core actions—"hand opening" and "fist clenching"—in the task design to reduce interference from multi-finger coupling and thus more accurately evaluate the real-time control performance of BMHI.

In further comparative experiments, we validated the core advantage of BMHI—its decoding strategy grounded in replicating the hierarchical transmission of the motor pathway. In the experimental group, BMHI decoded cortical activity to reconstruct muscle activation, which were then transmitted to myoelectric interfaces for motor execution. In contrast, the control group employed a conventional BHI that directly predicted finger forces from EEG signals. The results demonstrated that BMHI substantially outperformed BHI in both decoding accuracy and generalization capability. This improvement arises from BMHI's physiologically constrained decoding architecture, which integrates biological priors into the neural decoding process to guide signal transformation in a more interpretable manner. In essence, BMHI can be viewed as an AI-enabled, physiology-informed decoding model that introduces a biologically grounded intermediate layer between cortical signals and motor outputs. By embedding such physiological constraints, BMHI not only replicates the natural neuromuscular transmission process but also enhances the explainability, stability, and adaptability of brain–computer decoding — a direction consistent with the recent advances in interpretable artificial intelligence.

Building upon the validation results from offline experiments, we further evaluated the real-time control performance of BMHI in an online environment. Experimental results demonstrated that BMHI reliably enabled real-time operation of both brain-controlled virtual hands and physical robotic hands. In virtual hand tasks, the system exhibited high response stability and feedback consistency; while in robotic hand tasks, BMHI successfully achieved natural hand opening and fist clenching control, maintaining satisfactory temporal coordination even during complex grasping scenarios. These findings demonstrate that BMHI not only reproduces muscle activity at the signal level but also functionally substitutes for biological muscle output, enabling the reconstructed EMG signals to drive myoelectric interfaces and thereby achieve regression-based, continuous, and natural control.

## Conclusion

This study proposes a novel motor brain–computer interface framework—the Brain–Muscle–Hand Interface (BMHI)—that reconstructs the motor pathway. Unlike conventional brain–computer interfaces that directly map cortical EEG activity to external actions, BMHI decodes EEG signals to reconstruct muscle-level EMG activity, thereby restoring the intermediate neuromuscular layer that naturally links brain intention to motor execution. This physiologically inspired design re-establishes the natural flow of motor information that is bypassed in traditional decoding schemes. The reconstructed EMG signals can then be applied to a myoelectric interface to achieve precise and continuous motor control. By preserving the hierarchical organization and information transformation of the human motor system, BMHI realizes stable, adaptive, and natural neural control beyond the limitations of classification-based BCIs.

### Future research directions

The present work lays the essential foundation for this line of research, providing the technical and conceptual basis upon which future developments can be built. Building on this foundation, future research will further extend the BMHI framework to cross-subject and clinical scenarios by establishing personalized and adaptive brain–muscle modelling mechanisms, thereby enabling motor control for individuals with high-level paralysis or muscular atrophy without the need for actual electromyography input.

## METHODS

### Signal acquisition

In this study, a wearable 16-channel non-invasive EEG recorder, the g. Nautilus, was used to record EEG signals at a sampling frequency of 500Hz. Following the International 10-10 system, 16 channels covering the frontal and parietal regions of the sensorimotor area were selected for recording. For EMG signal acquisition, 12 channels were used to target the primary muscles involved in finger movement, with a sampling frequency of 1000Hz. Fig. 11 shows the distribution of muscles in the forearm, while Table 3 lists the function of each muscle. To capture the EMG activity of each muscle during gesture movements, the electrodes of the 12 channels were evenly arranged on the forearm near the elbow. The electrode arrangement is shown in Fig. 2. Additionally, force signals were collected using a pressure sensor at a sampling frequency of 6.6Hz.

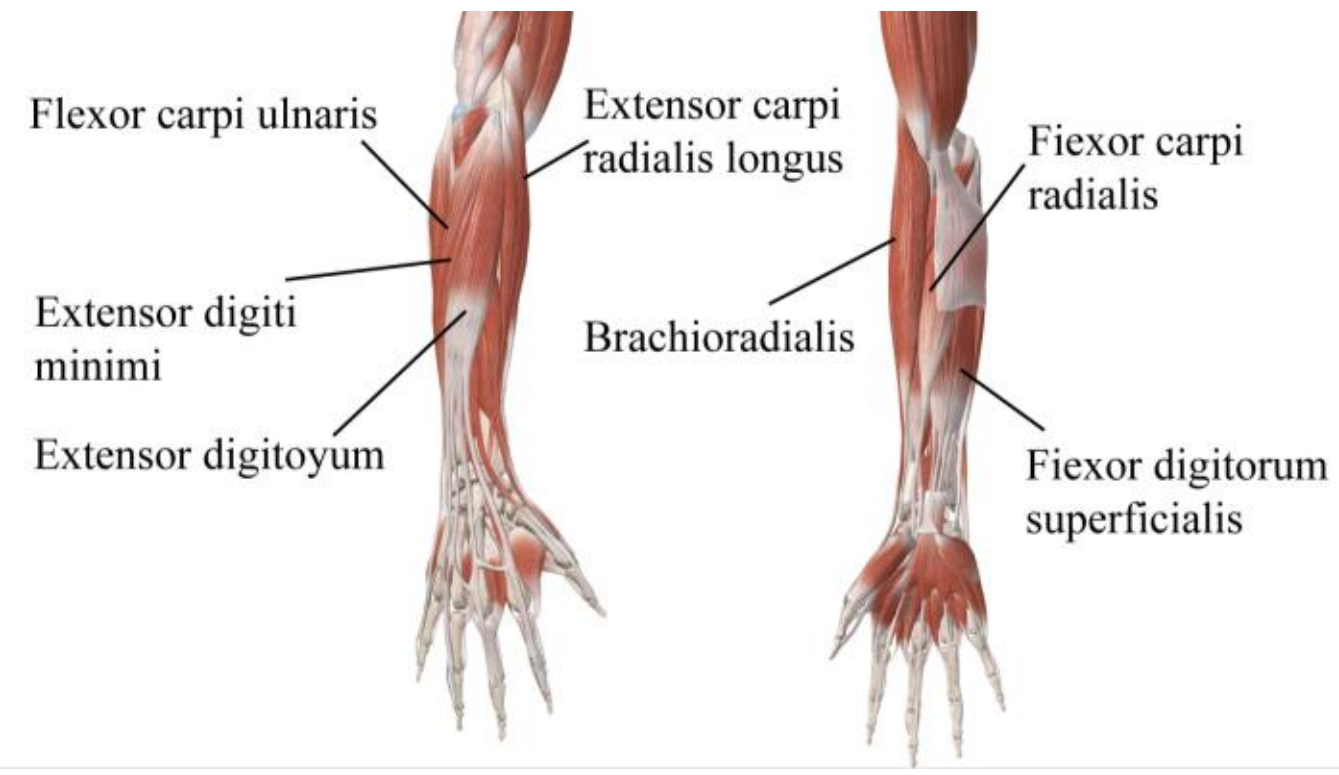

**Fig. 11. The distribution of muscles in the forearm**

**Table3. Small arm muscles and their role**

| Forearm Muscles | Main Functions |
|---|---|
| Brachioradialis | Elbow flexion |
| Flexor carpi ulnaris | Wrist flexion and ulnar deviation |
| Extensor degiti minimi | Fifth digit extension |
| Extensor digitorum | Finger extensor |
| Fiexor digitorum superficialis | Flexes the fingers |
| Flexor carpi radialis | Wrist flexion and radial deviation |
| Extensor carpi radialis longus | Wrist extension and radial deviation |

### Signal pre-processing

First, the EEG, EMG, and finger-force signals were temporally aligned by upsampling all signals to a uniform sampling rate of 1000Hz. Next, the EEG signals were processed using common average reference (CAR) and band-pass filtered between 15-35Hz. The CAR operation averages the signals from all channels to generate a reference value[34], which is then subtracted from each channel's signal to produce new channel signals. The mathematical expression for this operation is:

$$x_i^{CAR}(t) = x_i(t) - \frac{1}{C}\sum_{j=1}^{c} x_j(t)$$

where $x_j(t)$ is the potential value of the $j$ th channel and $C$ is the total number of channels. This processing step effectively removes common mode noise and enhances the spatial resolution of the signal.

For EMG signals, a 20-450Hz band-pass filter removed the DC offset while retaining muscle activity frequencies, followed by a 48-52Hz Butterworth notch filter to eliminate 50Hz interference, enhancing signal quality.

**Signal Segmentation**

The EEG, EMG, and finger forces were divided into 60 independent sessions. Each session analyzed hand signals to determine the subject's force generation state. Specifically, a threshold was set, and when the hand signal exceeded this threshold, the subject was considered to be performing the task. Signal segments meeting this criterion were extracted for further analysis. This segmentation ensures signal relevance and accuracy, providing high-quality task-related data for analysis.

**Brain-muscle-hand interface**

As illustrated in Figure 1, our proposed BMHI framework opens a new pathway for achieving natural motor control based on BCIs. The core concept of BMHI lies in replicating the neural motor pathway: it decodes EEG signals to reconstruct corresponding EMG signals, thereby functionally substituting biological muscles and further interfacing with myoelectric control systems. This mechanism allows BMHI to inherit the significant advantages of myoelectric interfaces in continuous control and movement fluidity, ultimately enabling natural and coherent motion control via regression models. The framework consists of two key stages: (1) the brain–muscle interface stage, which biomimetically replicates neural transmission from the brain to muscles by decoding EEG signals to reconstruct corresponding EMG signals; and (2) the muscle–hand interface stage, where the reconstructed EMG signals serve as control inputs to accurately estimate motor commands—such as finger force—based on established biomechanical mappings between muscle activation and hand motion. By constructing this hierarchical "brain–muscle–hand" information pathway, BMHI introduces a bio-inspired structure into signal decoding. This approach not only preserves the mature benefits of myoelectric interfaces in continuous control and movement naturalness, but also effectively mitigates common issues in traditional single-stage EEG-to-motion mapping methods, such as output instability and limited interpretability. Detailed technical implementation of the system will be elaborated in subsequent sections.

***Brain-muscle interface***

The top half of Fig. 12 shows the brain-muscle interface process. This paper proposes a sliding window-based CNN model to effectively capture EEG signal temporal and spatial features, providing representative input for brain-muscle interface signal mapping. We employ a convolutional neural network to model the complex nonlinear mapping between EEG signals and EMG features. CNN extracts spatial features through hierarchical convolutions, combined with a sliding window to capture local time-series information, gradually building a decoding pathway from brain to muscle signals. In order to let the model learn the coupling relationship between EEG signal and muscle signal and better predict the EMG signal, we designed a trend-based loss function whose mathematical expression is:

$$CovLoss(Y',Y) = \frac{1}{k}\sum_{i=1}^{n} cov(y_i',y_i)$$

where $Y$ is the label , $Y'$ is the model output, $y_i'$ represents the $i$th row of $Y'$, and $y_i$ represents the $i$th row of $Y$ .

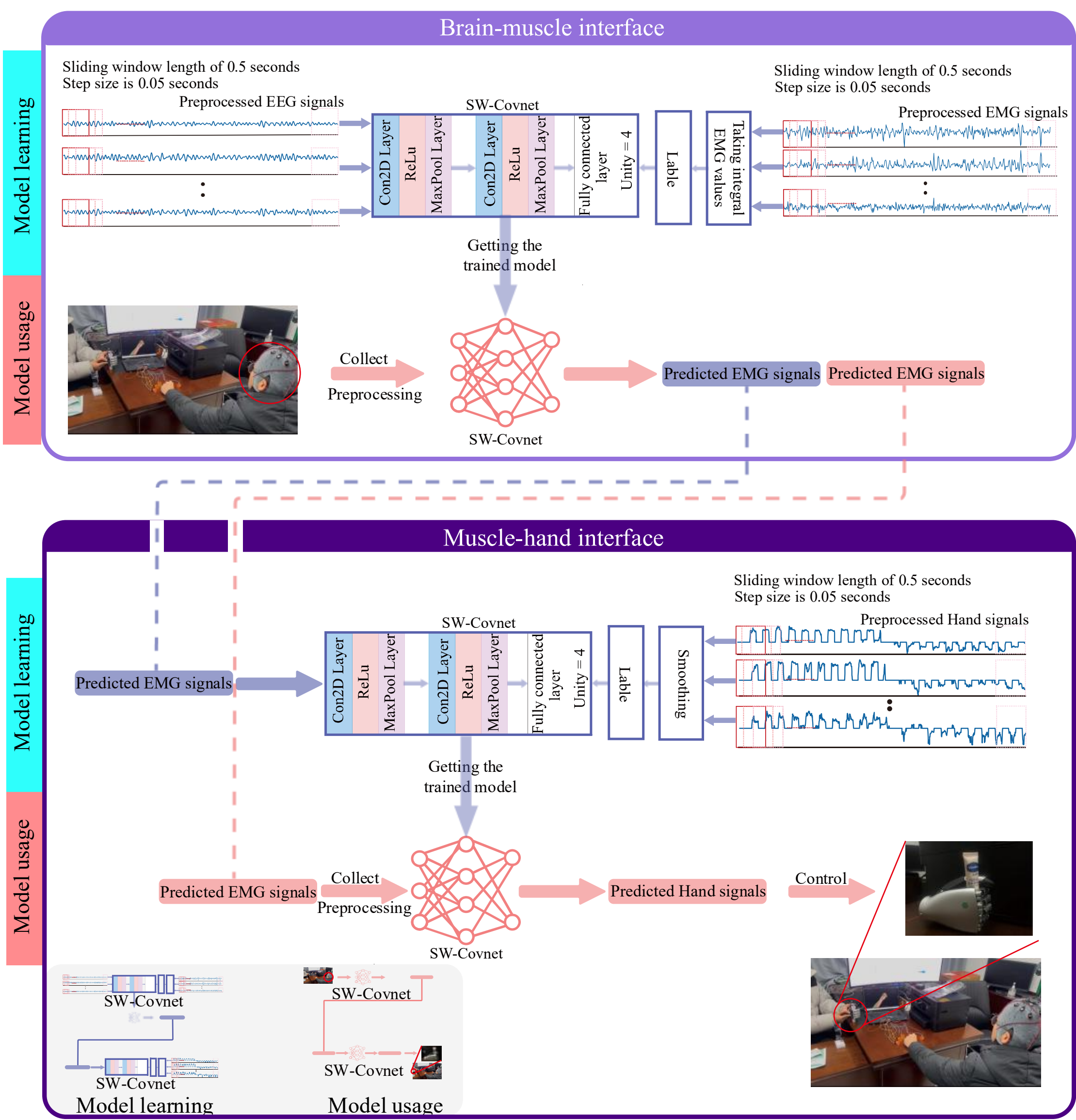

**Fig. 12. Brain-muscle interface flowchart**

***Muscle-hand interface:***

The lower part of Fig. 12 shows the process of the muscle-hand interface. To ensure that the EMG signals are accurately aligned with the samples of the hand signals, we introduce a sliding window approach from the brain-muscle interface processing to average the finger forces. In this way, the processed finger forces are able to be consistent in the time dimension with the reconstructed EMG signals. During the modeling of the muscle-hand interface, in order to capture the trend characteristics of the EMG signals more effectively, we further used the sliding window technique to smooth the EMG signals and hand signals separately. Specifically, the length of the sliding window was set to 50ms and the step size to 1ms to capture the trend of the signal at a fine-grained level. Subsequently, based on the smoothed signals, a convolutional neural network was used to model the mapping relationship between the EMG signal features and the finger forces.

## References and Notes

1. Shu T, Levine D, Yeon SH, Chun E, Shallal CC, McCullough J*, et al.* Tissue-integrated bionic knee restores versatile legged movement after amputation. **Science 2025**, 389(6756).
2. Ji Z, Yan J, Li CX, Sun YH, Wang SP, Tao J*, et al.* Precise rewiring of corticospinal axons and spinal interneurons via near-infrared optogenetics for spinal cord injury treatment. **Science Advances 2025**, 11(31).
3. Li W, Sheng RW, Cao MM, Rui YF. Exploring the Relationship Between Gut Microbiota and Sarcopenia Based on Gut-Muscle Axis. **Food Science & Nutrition 2024**, 12(11)**:** 8779-8792.
4. McDonald CL, Westcott-McCoy S, Weaver MR, Haagsma J, Kartin D. Global prevalence of traumatic non-fatal limb amputation. **Prosthetics and Orthotics International 2021**, 45(2)**:** 105-114.
5. Panichi LB, Stanca S, Dolciotti C, Bongioanni P. The Role of Oligodendrocytes in Neurodegenerative Diseases: Unwrapping the Layers. **International Journal of Molecular Sciences 2025**, 26(10).
6. Ijaz S, Mohammed I, Gholaminejhad M, Mokhtari T, Akbari M, Hassanzadeh G. Modulating Pro-inflammatory Cytokines, Tissue Damage Magnitude, and Motor Deficit in Spinal Cord Injury with Subventricular Zone-Derived Extracellular Vesicles. **Journal of Molecular Neuroscience 2020**, 70(3)**:** 458-466.
7. Resnik L, Huang H, Winslow A, Crouch DL, Zhang F, Wolk NJJon*, et al.* Evaluation of EMG pattern recognition for upper limb prosthesis control: a case study in comparison with direct myoelectric control. **2018**, 15(1)**:** 23.
8. Fleming A, Stafford N, Huang S, Hu X, Ferris DP, Huang HHJJone. Myoelectric control of robotic lower limb prostheses: a review of electromyography interfaces, control paradigms, challenges and future directions. **2021**, 18(4)**:** 041004.
9. Balbinot G, Li G, Wiest MJ, Pakosh M, Furlan JC, Kalsi-Ryan S*, et al.* Properties of the surface electromyogram following traumatic spinal cord injury: a scoping review. **2021**, 18(1)**:** 105.
10. Hochberg LR, Bacher D, Jarosiewicz B, Masse NY, Simeral JD, Vogel J*, et al.* Reach and grasp by people with tetraplegia using a neurally controlled robotic arm. **Nature 2012**, 485(7398)**:** 372-U121.
11. Ethier C, Oby ER, Bauman MJ, Miller LE. Restoration of grasp following paralysis through brain-controlled stimulation of muscles. **Nature 2012**, 485(7398)**:** 368-371.

12. Khan MA, Das R, Iversen HK, Puthusserypady S. Review on motor imagery based BCI systems for upper limb post-stroke neurorehabilitation: From designing to application. **Computers in Biology and Medicine 2020**, 123.
13. Khademi Z, Ebrahimi F, Kordy HM. A review of critical challenges in MI-BCI: From conventional to deep learning methods. **Journal of Neuroscience Methods 2023**, 383.
14. Roy AM. Adaptive transfer learning-based multiscale feature fused deep convolutional neural network for EEG MI multiclassification in brain-computer interface. **Engineering Applications of Artificial Intelligence 2022**, 116.
15. Cho H, Ahn M, Ahn S, Kwon M, Jun SC. EEG datasets for motor imagery brain-computer interface. **Gigascience 2017**, 6(7)**:** 1-8.
16. Lubin JA, Strebe JK, Kuo JS. Intracortical Microstimulation of Human Somatosensory Cortex Reproduces Touch in Spinal Cord Injury Patient. **Neurosurgery 2017**, 80(5)**:** N29-N30.
17. Lebedev MA, Nicolelis MAL. Brain-machine interfaces: past, present and future. **Trends in Neurosciences 2006**, 29(9)**:** 536-546.
18. Pruszynski JA, Kurtzer I, Nashed JY, Omrani M, Brouwer B, Scott SH. Primary motor cortex underlies multi-joint integration for fast feedback control. **Nature 2011**, 478(7369)**:** 387-+.
19. Nielsen JB. Human Spinal Motor Control. In: Hyman SE (ed). *Annual Review of Neuroscience, Vol 39*, vol. 39, 2016, pp 81-101.
20. Capaday C, Stein RJJoN. Amplitude modulation of the soleus H-reflex in the human during walking and standing. **1986**, 6(5)**:** 1308-1313.
21. Brunner C, Billinger M, Seeber M, Mullen TR, Makeig S. Volume Conduction Influences Scalp-Based Connectivity Estimates. **Frontiers in Computational Neuroscience 2016**, 10.
22. Onton J, Makeig SJPibr. Information-based modeling of event-related brain dynamics. **2006**, 159**:** 99-120.
23. Wolpaw JR, Acm. Brain-Computer Interfaces (BCIs) for Communication and Control. 9th International ACM SIGACCESS Conference on Computers and Accessibility; 2007 Oct 15-17 2007; Tempe, AZ; 2007. p. 1-2.
24. Saha S, Mamun KA, Ahmed K, Mostafa R, Naik GR, Darvishi S*, et al.* Progress in Brain Computer Interface: Challenges and Opportunities. **Frontiers in Systems Neuroscience 2021**, 15.
25. Wolpert DM, Diedrichsen J, Flanagan JRJNrn. Principles of sensorimotor learning. **2011**, 12(12)**:** 739-751.
26. Xu J, Mawase F, Schieber MHJPR. Evolution, biomechanics, and neurobiology converge to explain selective finger motor control. **2024**, 104(3)**:** 983-1020.
27. Schoffelen JM, Oostenveld R, Fries P. Neuronal coherence as a mechanism of effective corticospinal interaction. **Science 2005**, 308(5718)**:** 111-113.
28. Conway BA, Halliday DM, Farmer SF, Shahani U, Maas P, Weir AI*, et al.* Synchronization between motor cortex and spinal motoneuronal pool during the performance of a maintained motor task in man. **Journal of Physiology-London 1995**, 489(3)**:** 917-924.
29. Boonstra TW, Breakspear M. Neural mechanisms of intermuscular coherence: implications for the rectification of surface electromyography. **Journal of Neurophysiology 2012**, 107(3)**:** 796-807.

30. Bremner FD, Baker JR, Stephens JA. Variation in the degree of synchronization exhibited by motor units lying in different finger muscles in man. **The Journal of physiology 1991**, 432**:** 381-399.
31. Lang CE, Schieber MH. Human finger independence: Limitations due to passive mechanical coupling versus active neuromuscular control. **Journal of Neurophysiology 2004**, 92(5)**:** 2802-2810.
32. Kinoshita H, Kawai S, Ikuta K, Teraoka T. Individual finger forces acting on a grasped object during shaking actions. **Ergonomics 1996**, 39(2)**:** 243-256.
33. Zatsiorsky VM, Li Z-M, Latash MLJEbr. Enslaving effects in multi-finger force production. **2000**, 131(2)**:** 187-195.
34. Bertrand O, Perrin F, Pernier JJE, Section CNEP. A theoretical justification of the average reference in topographic evoked potential studies. **1985**, 62(6)**:** 462-464.

**Acknowledgments:** We thank our subjects for their contributions to this work.

**Funding:** This article is supported by the National Natural Science Foundation of China (62303423, 62373295,62103377), the STI 2030-Major Project (2022ZD0208500), Postdoctoral Science Foundation of China (2024T170844, 2023M733245), the Henan Province key research and development and promotion of special projects (242102311239, 252102311096), Shaanxi Provincial Key Research and Development Program (2023GXLH-012).

**Author contributions: Sun Ye:** Methodology, Validation, Visualization, Writing – original draft, Writing – review & editing. **Zuo Cuiming:** Data curation, Methodology, Software. **Zhang Rui:** Resources, Methodology, Writing – review & editing. **Shi Bin:** Data curation, Resources, Writing – review & editing. **Pang Yajing:** Methodology, Writing – review & editing. **Gao Lingyun:** Methodology, Writing – review & editing. **Zhao Bowei:** Formal analysis, Data curation. **Wang Jing:** Visualization, Resources, Writing – review & editing. **Yao Dezhong:** Project administration, Conceptualization, Writing – review & editing. **Liu Gang:** Project administration, Methodology, Validation, Visualization, Conceptualization, Formal analysis, Writing – review & editing, Resources.

**Competing interests:** The authors declare that they have no competing interests.

**Data and materials availability:** Data and materials are obtained by contacting corresponding authors.